\documentclass[graybox]{svmult}

\usepackage{type1cm}        
\usepackage{makeidx}         
\usepackage{graphicx}        
\usepackage{multicol}        
\usepackage[bottom]{footmisc}

\usepackage{newtxtext}       %
\usepackage{newtxmath}       

\makeindex             

\newcommand{\vsini}{$v$\,sin\,$i$}
\newcommand{\vmac}{$v_{\rm mac}$}
\newcommand{\Teff}{$T_{\rm eff}$}
\newcommand{\grav}{log\,$g$}
\newcommand{\kms}{km\,s$^{-1}$}
\newcommand{\vinfty}{$v_{\infty}$}
\newcommand{\mdot}{$\dot{M}$}
\newcommand{\micro}{$\xi_{\rm t}$}
\newcommand\ion[2]{#1\,{\sc #2}}

\begin{document}

\title*{A modern guide to \\ quantitative spectroscopy of massive OB stars}
\author{Sergio Sim\'on-D\'iaz}
\institute{Sergio Sim\'on-D\'iaz \at Instituto de Astrof\'isica de Canarias, c/ V\'ia L\'actea s/n, E-38200, La Laguna, Tenerife, Spain.\\
Departamento de Astrof\'isica, Universidad de La Laguna, E-38206, La Laguna, Tenerife, Spain.\\ \email{ssimon@iac.es}
}
%
%
\maketitle

\abstract{Quantitative spectroscopy is a powerful technique from which we can extract information about the physical properties and surface chemical composition of stars. In this chapter, I guide the reader through the main ideas required to get initiated in the learning process to become an expert in the application of state-of-the-art quantitative spectroscopic techniques to the study of massive OB stars.
\newline\indent}

\section{Introduction}
\label{sec:1}

Quantitative spectroscopy is one of the most rewarding fields of stellar astrophysics. It allows the courageous researcher who has decided to devote time to this fascinating enterprise, to have first hand access to an important set of empirical information about the investigated stars which, in most cases, cannot be obtained by any other means. This mainly comprises several key stellar parameters (including, e.g., spin rates, effective temperatures, and gravities), as well as surface abundances of those elements which have left their imprint in the observed piece of the stellar spectrum that will be analysed. In addition, the analysis process can help to highlight (and better characterize) the presence of stellar winds, circumstellar material, spots and surface magnetic fields, faint companions in binary systems, and/or some sources of stellar variability/activity, specially when multi-epoch spectroscopy is considered. 

I write {\em courageous researcher} because becoming a reliable expert on quantitative stellar spectroscopy requires mastering several skills comprising observational, theoretical, modeling and programming aspects, as well as having medium to high knowledge of radiative transfer and atomic physics. Also, from a practical point of view, my own expertise gives me confidence to remark that having a detail oriented profile certainly helps to avoid providing erroneous and/or spurious information from the analysed spectra.

But, what does quantitative spectroscopy means? Paraphrasing my good friend M.~A.~Urbaneja from the University of Innsbruck, quantitative spectroscopy can be defined in a simple way as the {\em inference of the physical parameters that (uniquely and completely?) characterize an astronomical object based on three tools: an observed spectrum, a set of theoretical spectra, and a given comparison metric}. 

Obviously, as the reader can imagine, this topic is so broad that it could lead to several books. In this chapter, I take the opportunity that the kind invitation\footnote{by D. Jones, J. Garc\'ia-Rojas and Petr Kabath (co-PI's of the {\em ERASMUS+ project ``Per aspera ad astra simul``})} to participate in this book offers me to provide a broad overview of the main quantitative spectroscopic techniques which are presently applied to the study of the so-called massive OB stars. 

The chapter is intended to serve to young students as a first approach to a field which has attracted my attention during the last 20 years. I should note that, despite its importance, at present, the number of real experts in the field around the world is limited to less than 50 people, and about one third of them are close to retirement. Hence, I consider that this is a good moment to write a summary text on the subject to serve as guideline for the next generations of students interested in joining the massive star crew.   

If you are one of them, please, use this chapter as a first working notebook. Do not stop here. Dig also, for further details, into the literature I quote along the text. And, once there, dig even deeper to find all the original sources explaining in more detail the physical and technical concepts that are presently incorporated into our modern (almost) automatized tools. 

Someone posed me the following question long time ago: {\em why a student needs to learn how to compute the square root of a 10-digit number if a calculator can easily do it?} If you know the answer to this question, I'm sure you can become one of the next experts in quantitative spectroscopy of massive OB stars. Go ahead and do a good job! You are really lucky to start in a fascinating time in which you will have easy access to thousands of observed and theoretical spectra, as well as to powerful computers allowing to run -- in a fast and efficient way -- analysis tools incorporating various types of comparison metrics. But, please, never forget the most important rule to enjoy what you are doing and make real progress: {\em don't use any of these nice tools as black boxes}.  

This chapter is structured in two main sections as follows. First, I put massive OB stars in context. Then, I describe the main tools and techniques presently used for quantitative spectroscopy of this important, but complex, group of stars. 

\section{Setting the scene: massive OB stars, from observations to empirical quantities}
\label{sec:2}

Quantitative stellar spectroscopy is an intricate tool which allow us to jump from observations to a set of empirical quantities defining a given star. Despite this is a general statement that can be applied to any type of star, the first thing one must realize is that the details of the intermediate steps defining a specific quantitative spectroscopic analysis -- as well as the outcome of such analysis -- depends on the domain of stellar parameters characterizing the star under study and the available piece of stellar spectrum. In this section, I describe the main ingredients and ideas that must be taken into account to understand the strengths and limitations of the main state-of-the-art tools and techniques used for the quantitative analysis of optical spectra of massive OB stars. 

\subsection{Massive OB stars in context}
\label{subsec:2.1}

The term OB stars is commonly used in the literature with several different (but related) meanings. Generally speaking, all of them refer to any given sample of stars with O and B spectral types
which define a specific group of interest to investigate a particular astrophysical question involving stars of this type. 

However, the considered range in spectral type, as well as the inclusion of luminosity classes other than dwarfs, varies from one study to another. For example, this term is used in some studies of stellar abundances in late-O and early-B dwarfs stars  (e.g., \cite{Daflon2004, Braganca2019}), but also in other works investigating the O and B star population of the Milky Way (e.g. \cite{Bouy2015, Mohr-Smith2017, Russeil2017}) or other galaxies in the Local Group (e.g., \cite{Garcia2014, Camacho2016, Oey2018}), or performing any quantitative empirical study of the physical properties of different subsamples including stars of this type (e.g., \cite{Degroote2010, Najarro2011, Przybilla2013, Wade2014, Fossati2015, SimonDiaz2018, Hennicker2018}). As a consequence, the range of mass and evolutionary status covered by the investigated sample of stars may differ. This fact can create some confusion between different communities and, hence, it is important to be highlighted from the very beginning. 

Along this chapter, I will follow the original definition by \cite{Morgan1951}, which identifies OB stars as a spectroscopic ''natural group`` which, at intermediate and high spectral resolution may be defined by the detection of helium lines in absorption. As indicated by N. R. Walborn in Chapter 3 of the book {\em Stellar spectral classification} by \cite{Gray2009}, the low temperature boundary of this group is a diagonal in the Hertzsprung-Russell diagram (HRD) running from B2\,V, through somewhat later types at intermediate luminosity classes, to the latest B\,Ia supergiants.

\begin{figure}[t!]
\begin{center}
\includegraphics[scale=.55]{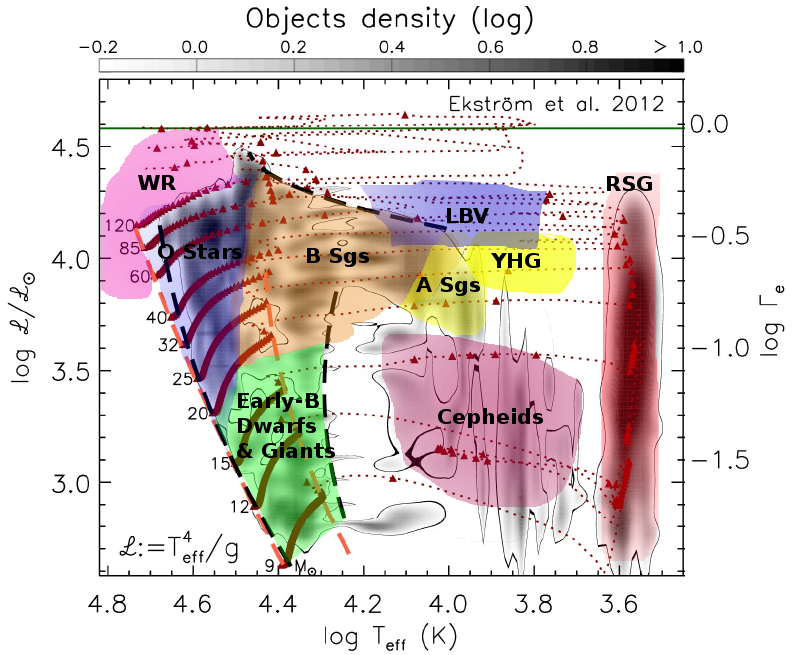}
\caption{Schematic representation of the realm of massive stars in the so-called spectroscopic Hertzsprung-Russell diagram (sHRD, \cite{Langer2014}). Colored regions depicts the approximate location of the various types of stellar objects (resulting from a phenomenological classification of their optical spectra) found to be associated with different evolutionary stages of stars born with masses above $\sim$9\,$M_{\odot}$. Figure is based on the plot from \cite{Castro2014} adapted by G. Holgado (PhD thesis, 2019). See \cite{Castro2014} for a detail descrition of the various lines and symbols depicted in the figure.}
\label{fig-1}       
\end{center}
\end{figure}

To put this group of stars in a broader context, I will use the schematic representation of the realm of massive stars in the so-called spectroscopic Hertzsprung-Rusell diagram\footnote{This diagram can be considered as an equivalent to the HRD, but only using stellar parameters derived spectroscopically (see also \cite{Langer2014}).} (sHRD) created by my former PhD student, G. Holgado, for his thesis \cite{Holgado2019}. Figure~\ref{fig-1} is an adaptation of a figure included in \cite{Castro2014}, where they presented for the first time the observational distribution of Galactic massive stars in the sHRD.  In its original form, \cite{Castro2014} presented a density map of stars in the uppermost part of the sHRD using a compilation of spectroscopically derived effective temperatures and gravities for almost 600 stars. Figure~\ref{fig-1} also includes, for reference purposes, the evolutionary tracks resulting from the non-rotating stellar evolution models for stars with masses in the range 9\,--\,120\,$M_{\odot}$ computed by \cite{Ekstrom2012}; and, overplotted as colored regions, the approximate location of various types of stellar objects associated with different evolutionary stages of massive stars\footnote{\cite{Poelarends2008} propose a fiducial value of 9\,$M_{\odot}$ for the minimum initial mass of massive stars at solar metallicity, where massive star is defined as a star that is massive enough to form a collapsing core at the end of its life and, thus, avoid the while dwarf fate (\cite{Langer2012}).}.

This figure illustrates how the original definition of OB stars by \cite{Morgan1951} does not only define a ''natural group`` from a spectral classification point of view, but also nicely covers the first part of the evolution of massive stars, fully including the main sequence (MS) as well as some early phases of the post-MS evolution\footnote{Actually, it is not yet completely clear whether the group of stars marked as B supergiants (B~Sgs) are post-MS stars, MS stars or, even, some of them are post red supergiant stars (see Section 6.1 in \cite{Langer2012} and references therein).}. 
The first thing to note is the broad range in mass ($M$\,$\approx$\,9\,--\,90\,$M_{\odot}$), effective temperature (\Teff\,$\approx$\,10000\,--\,55000\,K), and gravity (\grav\,=\,4.4\,--\,1.2\,dex) covered by OB stars. Although not represented in this diagram, these stars also cover a broad range in luminosities\footnote{Due to their high temperatures and luminosities, OB stars are sometimes also quoted as blue massive stars, and the O and B supergiants, as blue supergiants.} ($L$\,$\approx$\,10$^{\rm 3.5}$\,--\,10$^{\rm 6}$\,$L_{\odot}$) and projected rotation velocities (values of \vsini\ can reach up to 450\,\kms). They also develop radiatively driven winds (\cite{Kudritzki2000}), which become directly observable in their spectral energy distributions and spectral lines above $\approx$10$^{4}$\,$L_{\odot}$ (or, equivalentely, the 15\,$M_{\odot}$ track), and are mainly characterized by two global parameters: the terminal velocity (\vinfty) and the rate of mass loss (\mdot). All these extreme conditions must be taken into account when modelling the atmosheres of OB stars, a necessary step to perform any quantitative analysis of their spectra (see Section~\ref{subsec:3.2}).

From an evolutionary point of view, as mentioned above (see also Figure~\ref{fig-1}), OB stars represent the early evolutionary stages of massive star evolution, where early-B dwarfs/giants and O stars (including all luminosity classes) cover a different range in mass in the Main Sequence, while B~Sgs are the evolved descendants of the O-type stars. Other stellar objects associated with later phases of massive star evolution (depending on the initial mass) are the A~Sgs, the yellow hypergiants (YHG) and the luminous blue variables (LBV), and, last, the Wolf-Rayet stars (WR), the Red Supergiants (RSG) and the Cepheids. 

Although any deeper mention to massive star evolution is out of the scope of this chapter, I refer the interested reader to a recent review by N. Langer (\cite{Langer2012}) about pre-supernova evolution of massive single and binary stars as starting point. Most of the important references to learn further about this subject can be also found there.

\subsection{Why to care about massive OB stars?}

In addition to the interest per se within the field of stellar astrophysics, any in-depth study of massive OB stars is motivated by the huge impact that our knowledge of the basic physical properties and the evolution of these stars have on many and diverse aspects of the study of the Cosmos (e.g. star formation, chemodynamical evolution of galaxies, re-ionization of the Universe; see \cite{Elmegreen1977, Preibisch2007, Prantzos2008, TenorioTagle2006, Bromm2009, Robertson2010}). They are also the progenitors of the most extreme stellar objects  known in the Universe, some of them already quoted in previous section (e.g., hyper-energetic supernovae, Wolf-Rayet stars, luminous blue variables, massive black holes, neutron stars, magnetars, massive X and $\gamma$-ray binaries), and the origin of new studied phenomena such as long duration $\gamma$-ray bursters (\cite{Woosley2006}) or the recently detected gravitational waves produced by a merger of two massive black holes or neutron stars (\cite{Abbott2016, Abbott2017}; incl. LIGO and Virgo collaborations).

From a practical perspective, massive OB stars have become valuable indicators of present-day abundances in the Milky Way and other external galaxies, even beyond the Local Group (e.g., \cite{Urbaneja2005a, Urbaneja2005b, Castro2012}). In particular, they cannot only be considered as a reliable alternative to \ion{H}{ii} regions as abundance indicators, but also these stars are superior to nebulae in that they are not affected neither by depletion into dust grains, nor the long-standing problem of the discrepancy resulting from the computation of nebular abundances using collisional emission lines or recombination lines (see, e.g. \cite{Peimbert1967, Stasinska2002, GarciaRojas2007}).

In addition, in recent years, blue supergiants have been promoted to the hall of fame of the ''standard candles``, traditionally including cepheid and RR Lyrae variables, novae, Type Ia and Type II supernovae, as well as globular clusters and planetary nebulae (\cite{Alloin2003, Kudritzki2003}). Indeed, as highlighted by \cite{Kudritzki2012}, these stars are ideal stellar objects for the determination of extragalactic distances, in particular, because they are the brightest stars in the Universe and the perennial uncertainties troubling most of the other stellar distance indicators -- namely, interstellar extinction and metallicity -- do not affect them.

Last, the interpretation of the light emitted by close-by and distant \ion{H}{ii} regions and starburts galaxies relies on our knowledge of the effect that the strong ionizing radiation emitted by the O-type stars produce in the surrounding interstellar medium ({\cite{Stasinska1996, Stasinska1997, Leitherer1999}). Also, any empirical information extracted from the spectra of OB stars about spin rates, mass loss rates and wind terminal velocities, photospheric abundances, binarity, and/or stellar variability associated with any type of pulsational-type phenomena is of ultimate importance to step forward in our understanding of the evolution and final fate of massive stars (e.g. \cite{Lefever2007, Vink2010, Brott2011a, Brott2011b, Martins2013, Martins2013b, Castro2014, SimonDiaz2018, Markova2018, Pedersen2019, Bowman2019}), as well as of the chemodynamical impact that these extreme stellar objects have on the surrounding interstellar medium at different scales.

\subsection{Spectroscopy of massive OB stars}
\label{subsec:2.3}

\begin{figure}[t!]
\begin{center}
\includegraphics[width=0.39\textwidth, angle=90]{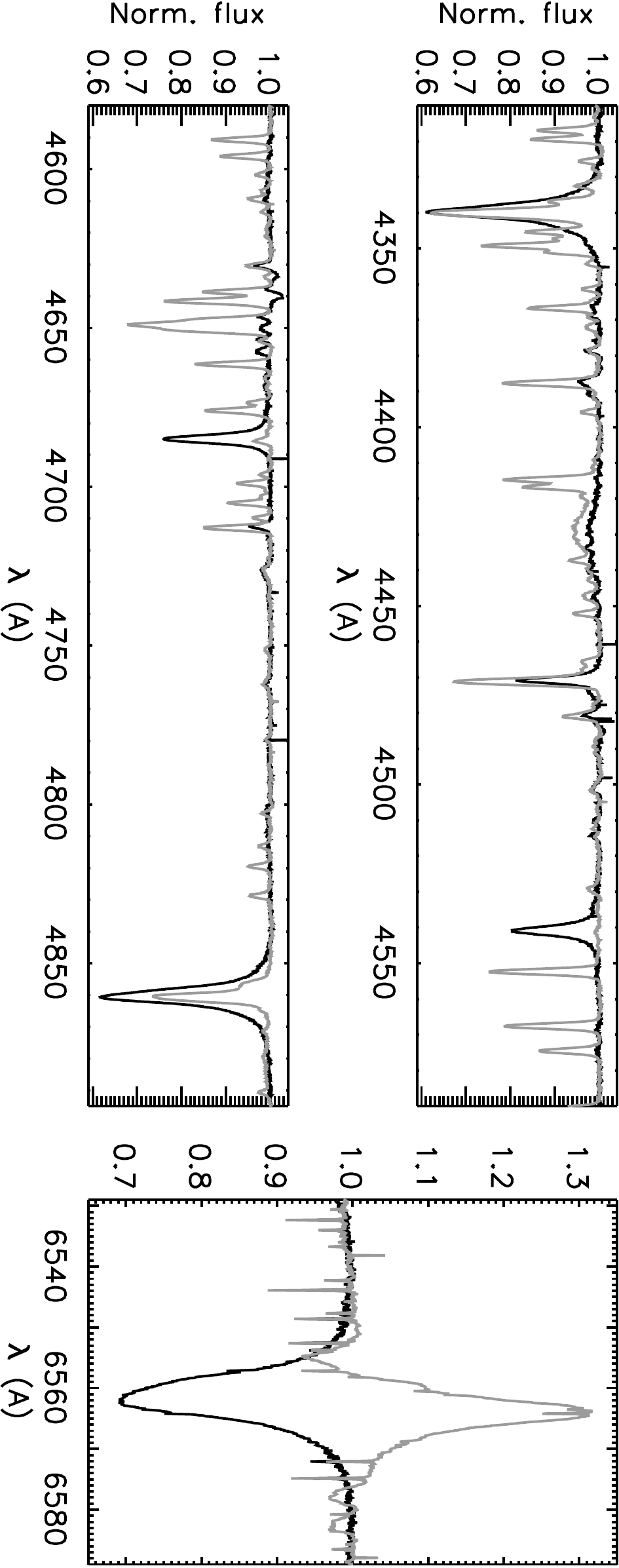}
\caption{High quality spectra of two Galactic OB stars obtained with the HERMES spectrograph attached to the Mercator1.2m telescope in the Roque de los Muchachos observatory (La Palma, Spain). Black and grey lines correspond to the mid-O dwarf HD~199579 and the early-B supergiant HD~2905, respectively. The differences between both spectra rely on the different physical properties and chemical composition of the outermost layers of each of the two stars. Quantitative spectroscopy is a powerful tool which allows to extract this information by comparing an observed spectrum with a grid of synthetic spectra obtained by means of a stellar atmosphere code. }
\label{fig-2}       
\end{center}
\end{figure}

There are three main spectral windows which are commonly considered to extract spectroscopic information about massive OB stars: 

\begin{itemize}
 \item (far-)UV: as provided by spectrographs on board the IUE, FUSE and HST space missions, covering some pieces of the 900\,--\,2200\,\AA\ spectral range;
 \item optical: covering either the full range between 3800 and 9000~\AA, or several key windows including the main set of diagnostic lines; and
 \item (near-)IR: mainly covering the H-, K- and L- bands at $\approx$1.62\,--\,1.77, 2.07\,--\,2.2, and 3.7\,--\,4.1$\mu$m, respectively.
\end{itemize}

Figure~\ref{fig-2} depicts three pieces of a typical optical spectrum of a mid-O dwarf (HD\,199579, black line) and an early-B supergiant (HD\,2905, grey line). These stars have been selected to illustrate how the different characteristics of stars in the OB star domain affect their spectra. For example, while in both cases the hydrogen Balmer lines (including H$_{\gamma}$, H$_{\beta}$, and  H$_{\alpha}$ at $\lambda\lambda$4341, 4860, and 6561\,\AA, respectively) are among the most prominent spectroscopic features, some particular characteristics of these diagnostic lines depend on the specific combination of surface gravity and effective temperature of the stars, as well as the existence of a more or less prominent stellar wind. For example, the larger the surface gravity for a given effective temperature, the more extended the wings of the Balmer lines; or, the stronger the wind density, the more remarkable the filling (in emission) of the H$_{\alpha}$ line. Also, while the presence of the He lines in absorption (including \ion{He}{i}$\lambda\lambda$4387, 4471, 4713\,\AA, and \ion{He}{ii}$\lambda\lambda$4541, 4686\,\AA) is the main identifier of an OB star, most of the \ion{He}{ii} lines disappear in the B-type stars. 

\begin{figure}[t!]
\begin{center}
\includegraphics[width=0.47\textwidth, angle=90]{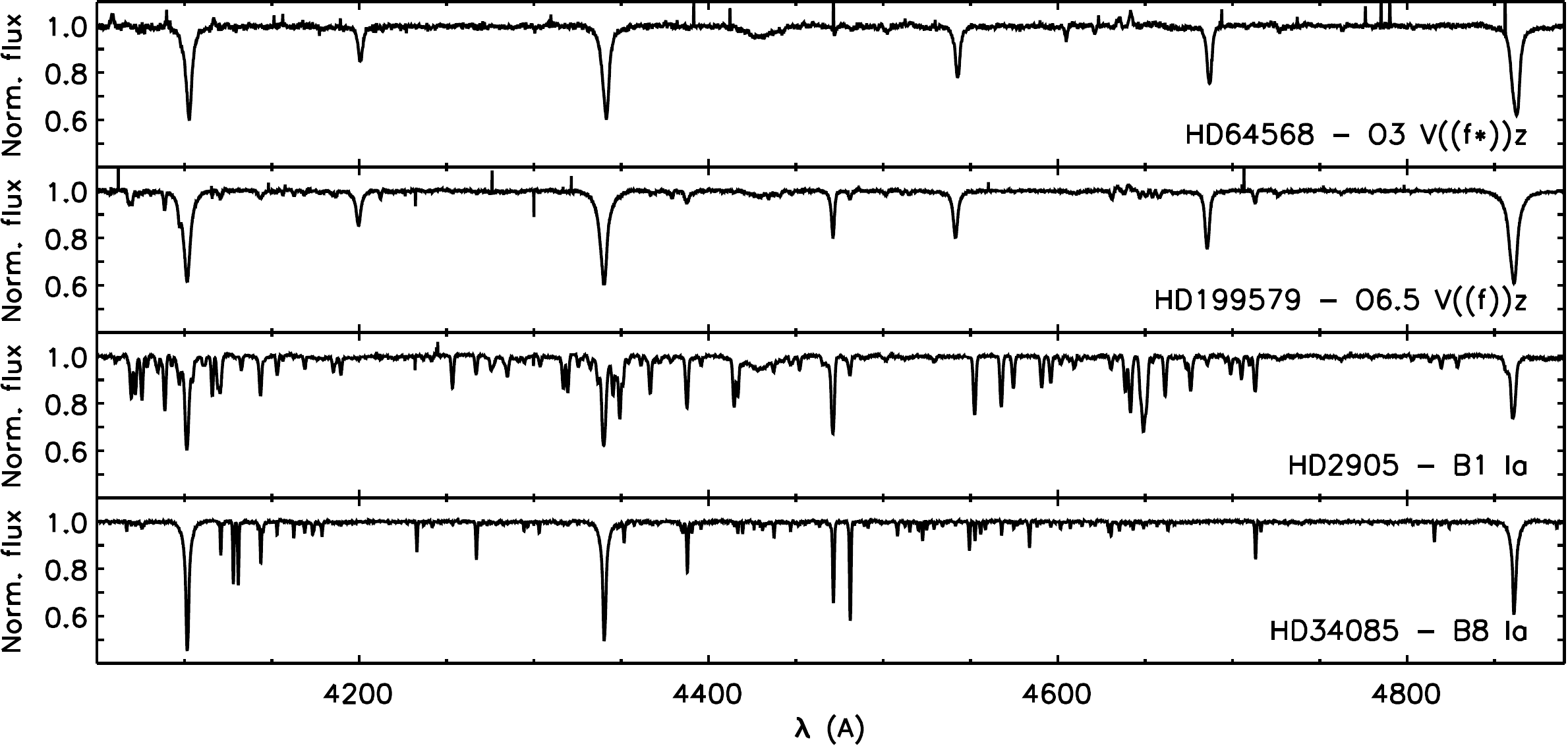}
\caption{From top to bottom, illustrative examples of optical spectra of an early-O, a mid-O, an early-B and a late-B star. Some representative lines of \ion{H}{i}, \ion{He}{i-ii}, \ion{O}{ii}, \ion{Si}{ii-iii-iv}, \ion{Mg}{ii}, \ion{N}{v} and \ion{C}{ii} are indicated in the corresponding spectrum where they are stronger.}
\label{fig-3}       
\end{center}
\end{figure}

The later has important consequences for the spectroscopic determination of effective temperatures, for which having access to lines from two consecutive ions is compulsory. In this sense, as will be further described in Sect.~\ref{subsec:3.4}, while the ratio of line strengths (or equivalent widths) of \ion{He}{ii} and \ion{He}{i} lines has traditionally been  considered as the main diagnostic to constrain the effective temperature of O4\,--\,O9.7 stars, other combination of diagnostic lines must be used in the B and early-O star domains\footnote{For example, \ion{N}{iii-iv-v} and \ion{Si}{iv-iii-ii} lines in the early-O and the early-B stars, respectively (see Section~\ref{subsec:3.4}).}, where either the \ion{He}{ii} or the \ion{He}{i} lines, respectively, disappear (Fig.~\ref{fig-3}). This fact creates a natural separation between early-O, mid/late-O, early-B, and mid/late-B stars in terms of the specificities of the quantitative spectroscopic analysis techniques to be applied. 

A similar situation occurs when dealing with lines of other elements beyond hydrogen and helium. Again, given the broad range in effective temperatures covered by OB stars, the number and strength of metal lines populating the optical spectra of, e.g. a mid-O dwarf, an early-B dwarf/supergiant, and a late-B supergiant varies a lot (see, again, Fig.~\ref{fig-3}). As a consequence, the potential estimation of surface abundances of the typical set of key elements that are normally considered in the study of OB stars (mainly He, C, N, O, Si, and Mg, but also Ne, S and Fe) must be based on lines from different ions depending on the effective temperature of the star under analysis\footnote{Also, different implementations of the associated model atoms -- including a more or less detailed description of the energy levels and transitions of specific ions -- are required.}. Indeed, in O-type stars and mid/late B Supergiants, the number of metals with available diagnostic lines in the optical spectrum is much more limited than in the early-B star domain, hence hampering the determination of the corresponding abundances. 

\begin{figure}[t!]
\begin{center}
\includegraphics[width=\textwidth]{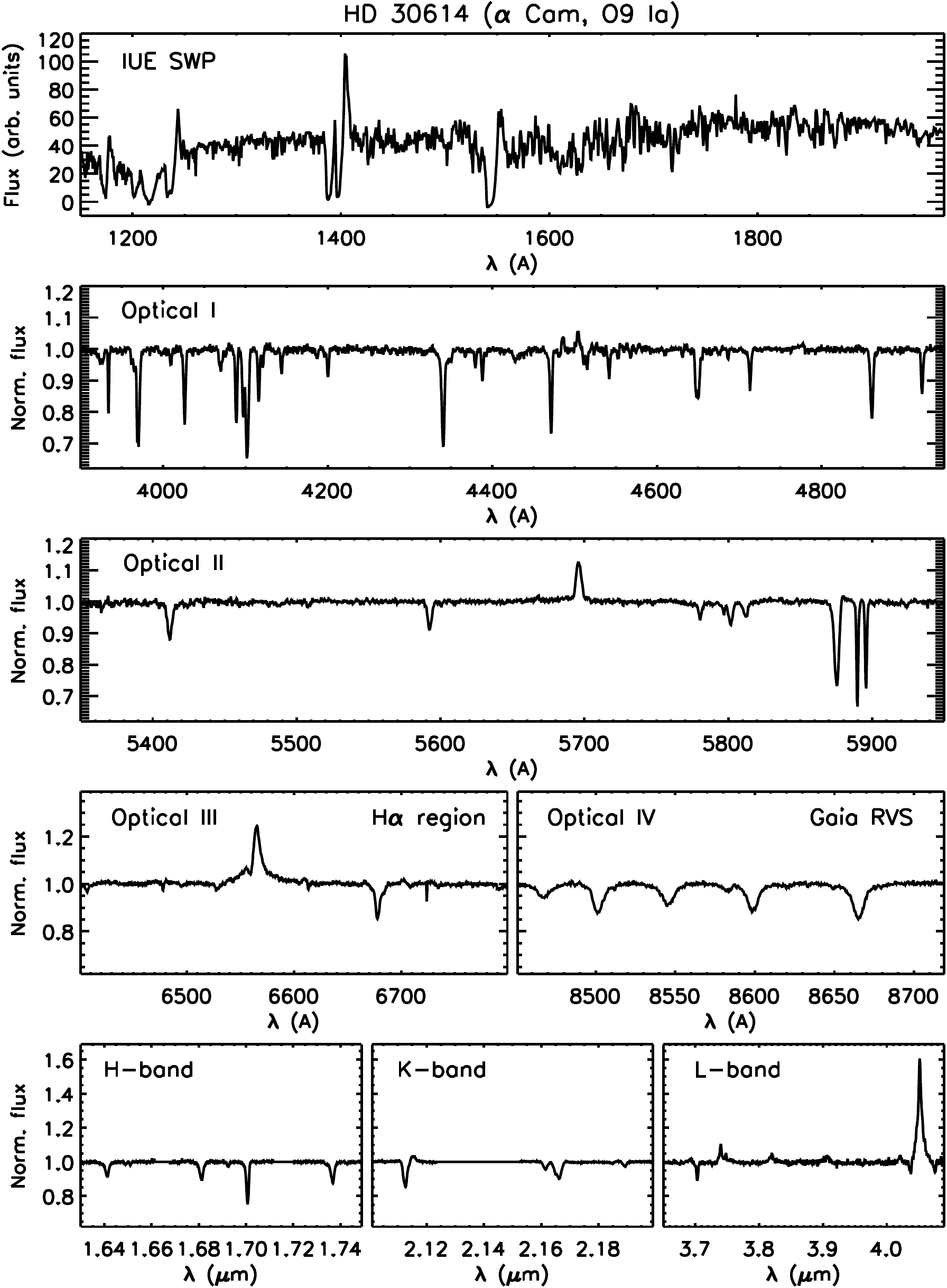}
\caption{Panchromatic view of the O9~Ia star HD~30614 ($\alpha$~Cam). The figure has been created combining observations obtained with different spectrographs attached to several ground based telescope facilities (optical and IR) and the IUE space mission (UV). Spectra kindly provided by F. Najarro \& M. Garcia (Centro de Astrobiolg\'ia, Madrid).}
\label{fig-4}       
\end{center}
\end{figure}

What about the other two spectral ranges? Despite the number of studies found in the literature performing quantitative spectroscopic analyses of OB stars in the optical range is much larger than those based on UV and/or IR spectra, the later contain important empirical information about the winds of these stars which is not directly accessible from the analysis of the optical spectrum (see below). In addition, they provide complementary information about effective temperatures, surface gravities, and abundances. Last, the quantitative spectroscopy in the (near)-IR range has been proposed as an important alternative to investigate the physical properties and chemical abundances in massive stars in highly obscured star forming regions (e.g. in the galactic center of the Milky Way), where the stars are much fainter in optical and UV wavelengths; hence, making more difficult to have access to high quality (mostly in terms of signal-to-noise ratio) spectra.

Next sections will be mainly devoted to the description of some of the tools and techniques presently used to perform quantitative spectroscopic analysis of OB stars based on their optical spectra. However, before entering into details, a few notes on these two other spectral ranges are worthwhile. The reader is also referred to the works performed by F. Najarro, M. Garcia, and T. Repolust (including a battery of papers and their PhD thesis), as well as \cite{Martins2011, Bouret2012, Lenorzer2004, Nieva2011}, and \cite{Marcolino2017}, among others, for further details on quantitative spectroscopic analysis performed in the UV and IR.

Figure~\ref{fig-4} provides a panchromatic view of the O9~Ia star HD~30614 ($\alpha$~Cam). The spectral windows depicted in Figures~\ref{fig-2} and \ref{fig-3} are now complemented with other portions of the spectrum of this star, including the UV range covered by the SWP spectrograph on board on the IUE satellite (top panel), the H, K and L bands in the IR (bottom panels), as well as two intermediate regions of the spectrum which have been elusively utilized for quantitative spectroscopy until recently. 

The Gaia RVS range (optical IV) has been included for completeness, and to illustrate how ''boring`` (and mostly useless) is this spectral range for the case of O- and early B-type stars\footnote{For this type of stars, the Gaia RVs range is basically populated by a few Paschen lines.}. This is not the case for the spectral window between 5400 and 5900\,\AA\ (optical II), in which it can be find very useful diagnostic lines for the study of O-type stars such as, e.g., \ion{He}{ii}$\lambda$5411\,\AA, \ion{O}{iii}$\lambda$5591\,\AA, \ion{C}{iii}$\lambda$5696\,\AA, \ion{C}{iv}$\lambda\lambda$5801, 5811~\AA, as well as one line that can be used as a powerful diagnostic to detect double line spectroscopic binaries: \ion{He}{i}$\lambda$5875\,\AA\ line.  

Regarding the UV and IR parts of the spectrum, I want to specially highlight the three P-Cygni profiles\footnote{Corresponding to the transitions \ion{N}{v}$\lambda\lambda$1239/43,\AA, \ion{Si}{iv}$\lambda\lambda$1394/403\,\AA, and \ion{C}{iv}$\lambda\lambda$1548/51\,\AA.} found in the range $\sim$1200\,--\,1600\,\AA, as well as some of the hydrogen and helium lines located in the IR (e.g., Br$_{\gamma}$, Pf$_{\gamma}$, and Br$_{\alpha}$ at 2.17, 3.74, and 4.05\,$\mu$m, respectively; \ion{He}{i}$\lambda\lambda$1.70, 2.11, 3.70\,$\mu$m; \ion{He}{ii}$\lambda\lambda$1,69, 2.19\,$\mu$m). Most of these lines, as well as other few spectroscopic features present in other regions of the UV serve as key diagnostics to obtain information about the main physical properties of the stellar wind developed by some OB stars, such as the terminal velocity, the mass loss rate, the clumping factor and the presence of shocks in the wind (see, e.g., references quoted above). Indeed, these lines are much more sensitive to all these factors than the three main wind diagnostic lines found in the optical (i.e., H$_{\alpha}$, \ion{He}{i}$\lambda$5875\,\AA, and \ion{He}{ii}$\lambda$4686\,\AA). As a consequence, any quantitative spectroscopic analysis of an OB star with an important stellar wind contribution should ideally consider the full UV+optical-IR range.

\newpage

\section{Tools and techniques used for quantitative spectroscopy of massive OB stars}
\label{sec:3}

This section is aimed at providing the reader a basic guide to the various steps that are commonly followed to perform the quantitative spectroscopic analysis of different types of OB stars based on their optical spectra\footnote{Most of the ideas presented along this section can be easily extrapolated to any quantitative spectroscopic analysis of the UV and IR spectral windows, with the only difference that other diagnostic lines, model atoms, and physical assumptions in the modeling of the stellar wind must be considered. Also some parameters and abundances may be more difficult (or even impossible in some cases) to be constrained just using the UV and/or IR part of the spectrum.}, as well as to the main presently available tools and techniques. Some further reading on the subject can be found below, separated by different type of analysis:

\begin{itemize}
 \item Line-broadening parameters in O and B stars: \cite{Ryans2002, SimonDiaz2007, SimonDiaz2014a, Braganca2012};
 \item Spectroscopic parameters in O stars: \cite{Herrero1992, Herrero2012, Repolust2004, Markova2004, Mokiem2005, SimonDiaz2011, SabinSanjulian2014, Martins2015a, RamirezAgudelo2017, Holgado2018};
 \item Abundances in O stars: \cite{Martins2015a, RiveroGonzalez2012a, RiveroGonzalez2012b, Grin2017, Carneiro2018, Carneiro2019};
 \item Spectroscopic parameters and abundances in late-O and early-B stars: \cite{Kilian1992, Hunter2007, Trundle2007, SimonDiaz2010, Nieva2012, Nieva2017,  Irrgang2014, Cazorla2017, Cazorla2017b, Dufton2018, Braganca2019}; 
 \item Spectroscopic parameters and abundances in B-Sgs: \cite{Urbaneja2005a, Urbaneja2008, Przybilla2006, Dufton2005, Markova2008, Lefever2010, Castro2012, McEvoy2015}.
 \end{itemize}
 I also recommend the reader to have a look to:
 \begin{itemize}
  \item Chapter 1 in the book {\em Oxygen in the Universe} by Stasi\`nska et al. (\cite{Stasinska2012}),
 \end{itemize}
 as well as to the interesting reviews on:
 \begin{itemize}
  \item {\em Winds from hot stars} by R.-P.~Kudritzki \& J.~Puls (\cite{Kudritzki2000}),
  \item {\em Modeling the atmospheres of massive stars} by J.~Puls (\cite{Puls2009}),
  \item {\em Parameters and winds of hot massive stars} by R.-P.~Kudritzki \& M.~A.~Urbaneja (\cite{Kudritzki2009})
  \item {\em Non-LTE Model Atom Construction} by N.~Przybilla (\cite{Przybilla2010}),
  \item {\em UV, optical and near-IR diagnostics on massive stars} by F.~Martins (\cite{Martins2011}), and
  \item {\em Highly accurate quantitative spectroscopy of massive stars in the Galaxy} by M.~F.~Nieva \& N.~Przybilla (\cite{Nieva2017}).
 \end{itemize}

\subsection{Quantitative spectroscopy of massive OB stars in a nutshell}
\label{subsec:3.1}

I have always considered that an efficient strategy to acquire new knowledge and skills about a given topic starts by having access to a quick, rough overview of the all the main points of the subject one wants to learn about. Then, once you have a more or less clear idea of where you want to go, you can come back -- in sequential order and in more detail -- to all those steps needed to fulfil your final objectives.    

Let's then apply this strategy to learn about quantitative spectroscopy of massive OB stars! I enumerate below the complete list of intermediate milestones one has to pursue to perform a complete quantitative spectroscopic analysis of an OB star: 

\begin{enumerate}
 \item Acquisition of the observed spectrum.
 \item Pre-processing of the spectrum, including a first qualitative visual assessment, the continuum normalization and the radial velocity correction.
 \item Determination of the line-broadening parameters. This is the basic step to have access to {\em projected rotational velocities}.
 \item Identification of the stellar atmosphere code and atomic models best suited for the analysis of the star under study.
 \item Creation of a grid of stellar atmosphere models, also including the corresponding synthetic spectra and equivalent widths for the main set of diagnostic lines needed for the specific analysis one wants to perform.
 \item Identification of the analysis strategy best suited to extract information from the observed spectrum of the star under study (e.g., spectral synthesis, use of equivalent widths).
 \item Determination of the main set of spectroscopic parameters accessible through the analysis of the observed piece of spectrum (e.g., basically the effective temperature and surface gravity, but also the microturbulence, the abundance of the diagnostic lines used to estimate the effective temperature, and the wind strength parameter). In this case we refer to {\em stellar parameters determination}.
 \item Determination of surface abundances of interest (among those elements with available diagnostic lines in the observed spectrum). This task is also called {\em chemical abundance analysis}.
\end{enumerate}
In addition to these eight points, a complete characterization of the main physical properties of the star requires another two steps which must incorporate some extra empirical information not directly accessible from the analysis of the optical spectrum, namely:

\begin{enumerate}
 \item the absolute magnitude of the star, in order to obtain estimates for the stellar luminosity, radius and mass. 
 \item the terminal velocity of the stellar wind, in order to obtain the mass loss rate. 
\end{enumerate}
And now, let's go back to the beginning and enter into more details. 

\subsection{Getting ready!}
\label{subsec:3.2}

We are entering in an era in which many of the new students (and a large fraction of the stellar community) will not have the necessity of preparing and undertaking any observing campaign to have access to all the spectroscopic observations required for their PhD studies. However, this does not mean that they should forget about learning (at least) some basic concepts of observational stellar spectroscopy.

Figures~\ref{fig-2} to \ref{fig-4} show illustrative examples of superb quality spectra (in terms of resolving power, S/N and wavelength coverage) of several types of OB stars. These are ideal spectra for a comprehensive and highly accurate quantitative spectroscopic analysis. However, in many situations it will not be possible to gather spectra with such a high quality and, hence, one will have to find a compromise between quality and number of stars with spectra good enough for the purposes of the study to be developed. For example, in some cases it can be more important to have access to high S/N spectra even sacrificing spectral resolution (e.g., when performing a chemical abundance analysis of extragalactic B~Sgs; \cite{Urbaneja2008}), but in other situations is more critical to gather high-resolution data even if the S/N is somewhat poorer (e.g., when dealing with measurements of projected rotational velocities in O stars and B-Sgs, or when obtaining stellar parameters in O stars with spectra contaminated by nebular emission from the associated \ion{H}{ii} region; \cite{SimonDiaz2007, SimonDiaz2014a, SabinSanjulian2014}).

It is also important for the beginnner to realize that, in many cases, the optical spectrum of an OB star may be contaminated with some other spectroscopic features which are not directly associated with the  star itself (from narrow interstellar lines and diffuse interstellar bands to telluric lines from the Earth atmosphere and/or nebular emission lines). In addition, since in most cases the starting point of a quantitative analysis is a normalized spectrum, the normalization process may have introduced spurious effects on some of the diagnostic lines (e.g., in the global shape of the wings of the Balmer lines which, as indicated in Sect.~\ref{subsec:3.6}, are the main diagnostics to constraint the surface gravity in OB stars). Last, specially if one wants to extract information about the radial velocity of the star -- either from a single snap~shot spectrum or a time~series --, it is important to check whether the spectrum has been corrected from heliocentric/barycentric velocity; and, if not, learn how to do it.

All these questions will definitely affect the scope, accuracy, and reliability of any type of quantitative spectroscopic analysis, as well as its outcome. So, my first two advises before going ahead are (1) do not forget to {\em incorporate to your list of learnt skills} the main technical concepts about observational spectroscopy, and (2) do not start the quantitative spectroscopic analysis before performing a {\em qualitative (visual) assessment} of the observed spectrum to understand what you have in your hands. These two initial steps certainly help to establish the best strategy to follow, as well as to avoid over-interpretations of the outcome of the analysis. For example, a double line spectroscopic binary cannot be analysed in the same way as an isolated star, or the results for a chemical abundance analysis of an early-B star may be erroneous if one does not realize that is dealing with a Be star with a circumstellar emitting disc. 

\subsection{Radial velocity correction}
\label{subsec:3.3}

Even if the radial velocity of the star is not a piece of information required by the study one wants to develop, the observed spectrum must be corrected by Doppler shift to ensure that all diagnostic lines are located in the laboratory position. While this correction is not very critical in those parts of the analysis based on equivalent widths, it may have non-negligible consequences in the determination of spectroscopic parameters by means of any type of line-profile fitting technique. The later is specially critical when establishing the surface gravity in the case of O- and B-type stars, since this is based in the fitting of the wings of the hydrogen Balmer lines.

There are several standard techniques that can be applied to perform the radial velocity correction, including, e.g., identification of the core of one or several diagnostic lines, either visually or using a gaussian fit to the line profile, and/or cross-correlation with a template. The most important warning to take into account when dealing with OB stars, however, is that one must remember that some lines may be affected by stellar winds (e.g. \ion{He}{i}$\lambda$5875\,\AA\ and \ion{He}{ii}$\lambda$4686\,\AA; or even some metal lines in cases of stars with very strong winds), and hence the use of these lines may led to erroneous results. In addition, the usual techniques based in the identification of the core of the line may fail in those stars with a high projected rotational velocity or important asymmetries due to stellar oscillations.

\subsection{Line-broadening parameters}
\label{subsec:3.4}

One of the most straightforward and cheapest ways (from an observational point of view) to obtain information about stellar spins is based on the effect that rotation produces on the spectral lines: stellar rotation broadens the spectral lines. However, this is not the only line-broadening mechanism acting in O- and B-type stars. As reviewed by A. Herrero (2019) in Sect. 2.13 of the book {\em Radiative transfer in stellar and planetary atmospheres}, there are, at least, another five mechanisms to be taken into account in these hot, massive stars: the natural, thermal, collisional (mainly Stark, both linear\footnote{Linear Stark broadening mainly affect the wings of the H and, to a less extent, the \ion{He}{ii} lines; indeed, this effect is mainly used to constrain the surface gravity (see Sect.~\ref{subsec:3.6}).} and quadratic\footnote{Quadratic Stark broadening, which is much less pronounced than the linear one, mainly affects the shape of the \ion{He}{i} lines.}), microturbulent, and macroturbulent (pulsational?) broadenings, respectively. Therefore, the first step of the quantitative spectroscopic analysis consists in inferring the projected\footnote{into the line-of-sight} component of the equatorial rotational velocity (\vsini) by disentangling the effect that rotation produces on the line-profile from any other comparable effect produced by the remaining broadening mechanism. And, as the reader can imagine, an important part of the process will be the selection of the best suited lines for the line broadening analysis, taking into account that the less affected the diagnostic line by the other (non-rotational) broadenings or by blends with other lines, the better. For example, it is always better to use a well isolated photospheric metal line than a hydrogen or helium line.

From a technical point of view, as in those other steps of the quantitative analysis process in which we are extracting information from the shape of the line-profile ({\em vs.} use of equivalent widths), the resolving power of the observed spectrum (as well as the number of points defining the line) is one important limiting factor of the accuracy we can reach in the determination of \vsini. For example, as a general rule of thumb, a spectral resolution $R$ implies a rough minimum limit in a reliable determination of \vsini\ of $c$/$R$, where $c$ is the speed of light in the same units as \vsini. On the other hand, when the Fourier transform method is used to estimate the projected rotational velocity, due to the Nyquist theorem, the spectral dispersion ($\Delta\lambda$, in \AA/pix) of the stellar spectrum imposes a limit in the lowest \vsini\ that can be derived, roughly given by 1.320 $c \Delta\lambda$/$\lambda$ (see \cite{SimonDiaz2007}).

I refer the reader to \cite{SimonDiaz2007, SimonDiaz2014a} and references therein for a thorough description of the various methods that have been routinely applied in the last 60 years for the determination of projected rotational velocities in OB stars, also including a discussion of the pros, cons and limitations of each method.

\begin{figure}[t!]
\begin{center}
\includegraphics[width=0.46\textwidth, angle=90]{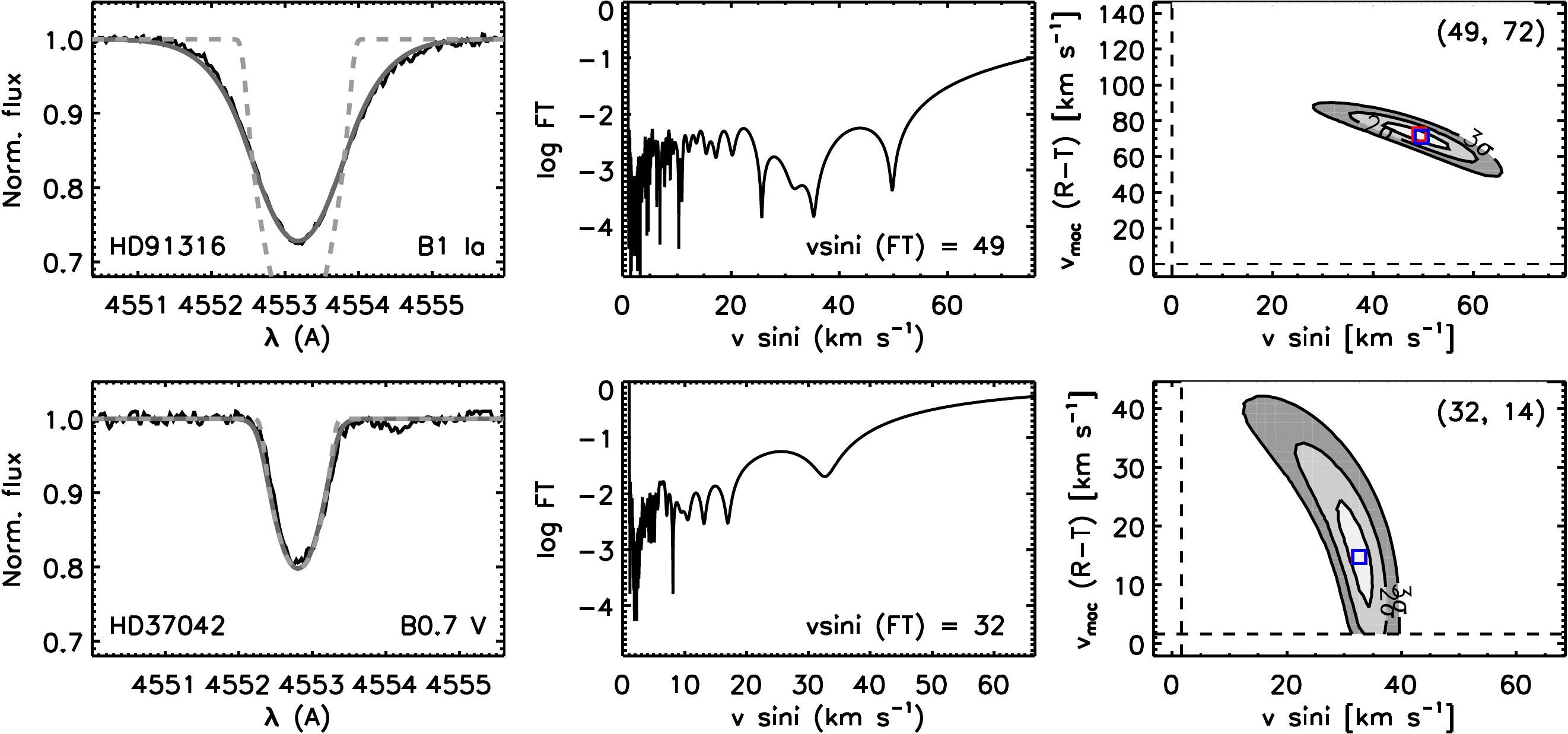}
\caption{Combined FT+GOF line-broadening analysis of the \ion{Si}{iii}$\lambda$4552\,\AA\ line in the early-B dwarf HD~37042 (bottom) and the early-B supergiant HD~91316 (top). [Left panels] The best fitting synthetic profile (solid gray) and the profile corresponding to \vsini(FT) and \vmac\,=\,0 (dashed gray) are over~plotted to the observed profile (solid black). [Middle panels] Fourier transform of the observed profile. [Right panels] $\chi^2$-2D-map resulting from the GOF analysis.}
\label{fig-5}       
\end{center}
\end{figure}

At present, the combined use of the Fourier transform (FT) and a goodness-of-fit (GOF) methods (Figure~\ref{fig-5}) has become a standard strategy to disentangle the effect of rotation from the other main sources of broadening shaping the line-profiles of OB stars. In brief, and following Gray (\cite{Gray1976}; see also the latest edition of the book, published in 2005), the Fourier transform method for the determination of \vsini\ is based on the intrinsic property of the rotational broadening function which develops zeroes in its Fourier transform. As firstly described by \cite{Carroll1933}, the position of these zeroes in frequency space depends on the \vsini\ of the star, so that the frequency of the first zero ($\sigma_1$) is related to the rotational velocity through:
\begin{equation}\label{eq:1}
 \frac{\lambda_0}{c}~v\,{\rm sin}i~\sigma_1\,=\,A
\end{equation}
where $\lambda_0$ is position in wavelength of the core of the line-profile, and $A$ is a constant that depends on the limb-darkening coefficient\footnote{The most common value used for $A$ is 0.660, which corresponds to a limb-darkening coefficient of $\epsilon$\,=\,0.6 (see, however, Fig.~3 in \cite{Aerts2014}).}.

Since the FT method only provides an estimate of \vsini, but its clear that this is not the only line-broadening agent even in the case of phostospheric metal lines (see, e.g. the case of the early B-Sg HD~91316 in Fig.~\ref{fig-5}), it needs to be complemented with a goodness-of-fit method, in which a $\chi^2$ fitting strategy if followed. In the later, an intrinsic profile\footnote{with the same equivalent width as the observed profile} -- which can be a $\delta$-function or a synthetic line resulting from a stellar atmosphere code --  is convolved with a rotational and a macroturbulent profile, and both line-broadening parameters (\vsini\ and \vmac) are obtained from the minimum value of the $\chi^2$-2D-map resulting from this GOF analysis.

The combined FT+GOF method provides a powerful and straightforward strategy to have access to the projected rotational velocity of a given star. In an ideal case, both determinations of \vsini, as resulting from the FT and the GOF methods should be in agreement. However, this is not always the case, implying that the situation for this specific star is more complex than initially expected due to the presence on the line profile of effects originated by, e.g., some types of stellar oscillations or spots and/or chemical inhomogeneities in the stellar surface \cite{Aerts2014}.

Overall, the FT method has been recently proven (\cite{SimonDiaz2007}) to be a better suited strategy to obtain actual estimates of projected rotational velocities in the whole OB star domain than other previously considered methods (\cite{Slettebak1975, Conti1977, Penny1996, Howarth1997, Ryans2002}). However, some limitations and caveats to be further investigated still remain (e.g., \cite{SimonDiaz2014a, SimonDiaz2017, Aerts2014, Sundqvist2013b}). This open new interesting lines of research for the new generation of massive star spectroscopists.

\subsection{Stellar atmosphere codes}
\label{subsec:3.5}

Stellar atmosphere codes are unavoidable tools when dealing with quantitative stellar spectroscopy. They are one of the major outcomes from an important, but complex research field which has led to a large number of texts in the literature. 

Although I assume that the reader of this chapter has acquired basic knowledge on radiative transfer in stellar atmospheres as part of her/his university studies, I encourage any young student willing to devote his/her career to quantitative spectroscopy of massive OB stars to deepen further into the specificities of stellar atmosphere modeling of hot stars already in the early stages of her/his career. The most recent version of the book {\em Theory of Stellar Atmospheres} by D.~Mihalas is a perfect starting point. I also recommend Chapter 4 of the book {\em Radiative transfer in stellar and planetary atmopsheres}, written by J. Puls, for a more in-depth description of the methods developed for the modeling of expanding atmospheres of early-type stars.\\

Before providing a brief overview of the main available stellar atmosphere codes for quantitative spectroscopy of OB stars, I quote below a few basic ideas allowing the non-expert in the field to easily understand the importance of model atmospheres for quantitative spectroscopy (see the sketch presented Fig.~\ref{fig-6}):
\begin{itemize}
 \item A stellar atmosphere is a thin layer in the surface of a star which does not have its own energy sources. Only redistribution of radiative energy takes place.
 \item A (one-dimensional, 1D) model atmosphere is a large table which describes the temperatures, pressures and many other properties of the gas as they vary with depth below the stellar surface.
 \item A stellar atmosphere is the part of the star where the emergent spectral energy distribution (including the continuum and the line spectrum) is formed.  
 \item A spectrum synthesis code is a computational tool that allows the calculation of the stellar emergent (synthetic) spectrum from a given model atmosphere.
 \item Quantitative stellar spectroscopy allows to extract information about the physical properties and chemical composition of a stellar atmosphere from the comparison of an observed spectrum and a grid of spectra computed with a spectrum synthesis code coupled to a stellar atmosphere code\footnote{In many cases, a stellar atmosphere code includes the computation of the emergent spectrum.}.  
\end{itemize}

\begin{figure}[t!]
\begin{center}
\includegraphics[width=\textwidth]{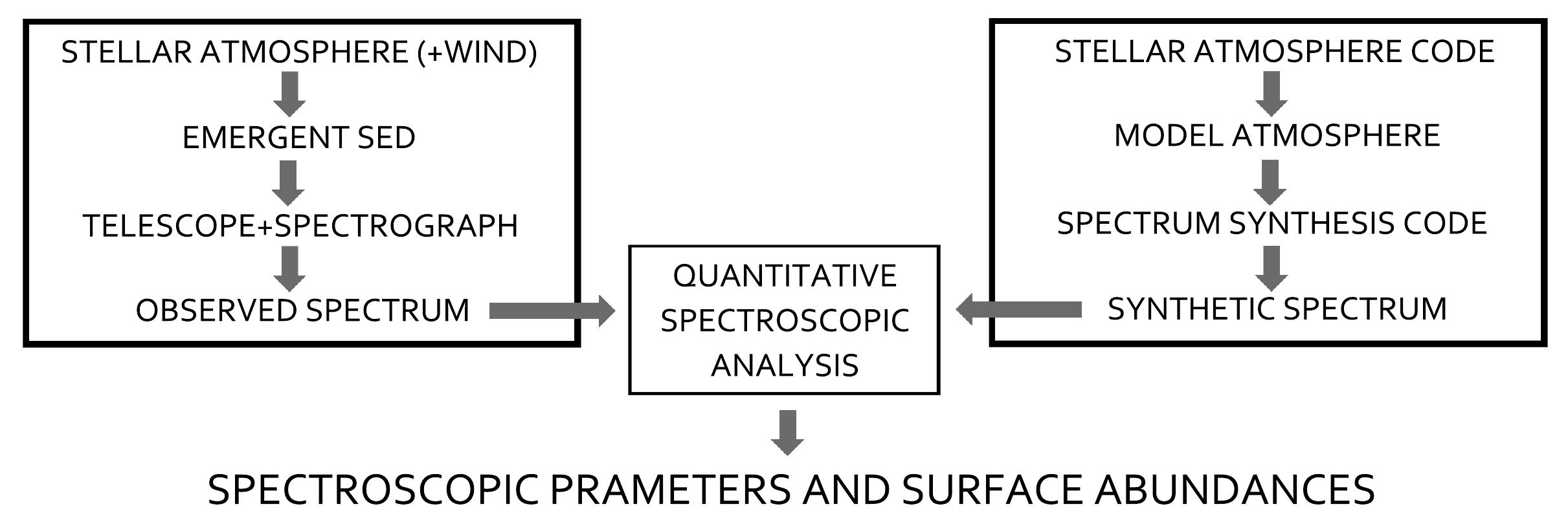}
\caption{Schematic representation of the main actors of quantitative stellar spectroscopy.}
\label{fig-6}       
\end{center}
\end{figure}

\subsubsection{State-of-the-art stellar atmosphere codes for OB stars}
\label{subsec:3.5.1}

The last decades of the XX century witnessed an enormous progress in the development of adequate stellar atmosphere codes for hot massive stars. The great efforts devoted by a small group of experts in the field, based on the firm theoretical foundations on radiative transfer laid by e.g. V.~V.~Sobolev, D.~Mihalas, L.~H.~Auer, L.~B.~Lucy, P.~M.~Solomon, J.~I.~Castor, D.~G.~Hummer, H.~J.~Lamers, J.~P.~Cassinelli (among others), have made it possible the massive star community to have access to a modern generation of stellar atmosphere codes which are allowing to perform reliable quantitative spectroscopic analyses of medium to large samples of O- and B-type stars in a reasonable amount of time.

This ambitious enterprise implied the inclusion of a realistic description of physical processes occurring in the outer layers of these extreme stellar objects such as, e.g., departure from the local thermodynamic equilibrium (non-LTE), line-blanketing and, in some cases, line-driven stellar winds. In particular, the later required the consideration of geometries departing from the simple plane-parallel approach, as well as the development of intricate computational techniques to deal with radiative transfer in (rapidly) expanding atmospheres (i.e. stellar winds).\\
\newline
At present, the stellar atmosphere codes most commonly used for the quantitative spectroscopic analysis of OB stars are:

\begin{itemize}
 \item  {\sc atlas} (\cite{Kurucz1970}) coupled with {\sc detail/surface} (\cite{Giddings1981}, Buttler \& Giddings 1985)
 \item  {\sc tlusty} (\cite{Hubeny1988}), coupled with {\sc synspec} (\cite{Hubeny2011})
 \item  {\sc cmfgen} (\cite{Hillier1998})
 \item  {\sc fastwind} (\cite{SantolayaRey1997, Puls2005, Puls2017})
 \item  {\sc p}o{\sc wr} (\cite{Grafener2002, Hamann2004, Sander2015})
 \item  {\sc wm}-{\em basic} (\cite{Pauldrach2001})
\end{itemize}

While all of them are 1-D, non-LTE, line-blanketed codes, they differ in how they treat geometry (plane parallel/spherical), hydrostatic equilibrium/mass outflows, line blanketing/blocking, micro- and macro-clumping ({\em vs.} unclumped winds), as well as the considered strategy to solve the complex, interwined set of equations of radiative transfer, and how they deal with information about atomic data\footnote{Model atoms -- including information about energy levels and the main collisional and radiative transitions between levels and/or the continuum -- are a very important ingredient of stellar atmosphere code. They will be only occasionally mentioned along this chapter; however, basic knowledge of how models atoms are implemented and used in stellar atmosphere and diagnostic codes is the forth pillar a quantitative stellar spectroscopist should dominate, along with basic concepts of observational stellar spectroscopy, radiative transfer and stellar atmosphere modeling.}.
As a consequence, not all these codes are equally optimized to analyse different types of OB stars or specific windows of the stellar spectrum. For example, both {\sc tlusty} and {\sc detail/surface} calculate occupation numbers/spectra on top of hydrostatic, plane parallel atmospheres; hence, they are ''only`` suited for the analysis of stars with negligible winds. Also, despite {\sc cmfgen}, {\sc fastwind}, {\sc p}o{\sc wr} and {\sc wm}-{\em basic} can, all of them, deal with spherically extended atmosphered with winds, the later ({\sc wm}-{\em basic}) is mainly applicable to the analysis of the UV range. Last, due to the different approximation considered by these codes for the treatment o line blanketing/blocking, the amount of computational time required varies from one code to other, ranging from less than one hour in the case of {\sc fastwind} and {\sc detail/surface} to several hours for {\sc cmfgen}, {\sc p}o{\sc wr}, {\sc tlusty} and {\sc wm}-{\em basic} models.

Further notes on state-of-the-art approaches to model the atmospheres of hot, massive stars, as well as improvements occurred in this field in the last years can be found in \cite{Puls2009} and \cite{Puls2015}, respectively. In particular, Table~1 in \cite{Puls2009} provides a nice overview of the main characteristics and range of applicability of all the stellar atmosphere codes quoted above.

\subsubsection{Grids of models for quantitative spectroscopy of OB stars}
\label{subsec:3.5.2}

At this point, we are almost ready to proceed with the determination of those stellar parameters that can be directly obtained through the spectroscopic analysis (see Section~\ref{subsec:3.6}). But, before, we need to spend a few time on the design and computation of a grid of stellar atmosphere models. Or, in same cases, we will be able to use directly any of the pre-computed grids which the developers (or their direct collaborators) have made publicly available. Some examples of the later can be found in the webpages of {\sc tlusty}\footnote{nova.astro.umd.edu}, 
{\sc cmfgen}\footnote{http://kookaburra.phyast.pitt.edu/hillier/web/CMFGEN.htm}, or
{\sc p}o{\sc wr}\footnote{http://www.astro.physik.uni-potsdam.de/~wrh/PoWR/powrgrid1.php}. However, these are not the only available grid of models; many others have not been done public, but could be available with permission of the owners. This is, e.g. the case of the vast grid of {\sc fastwind} models covering the O star domain (for solar and half solar metallicity) computed at the Instituto de Astrof\'isica de Canarias and which is presently incorporated to the {\sc iacob} grid based automatized tool ({\sc iacob-gbat}, see \cite{SimonDiaz2011, SabinSanjulian2014, Holgado2018}).

Regardless of using a pre-computed grid, or creating a new one, there are a few key points that must be carefully checked before performing the quantitative spectroscopic analysis:
\begin{itemize}
 \item The stellar atmosphere code used to compute the grid must consider all important physical processes occurring in the star under study (see Section~\ref{subsec:3.5.1}). The same consideration must be taken into account for the other geometrical and dynamical aspects of the modeling.
 \item All the key diagnostic lines must be properly included and treated in the computation of the associated synthetic spectra. Obviously, the wavelength coverage of the grid of synthetic spectra include the observed spectrum.
 \item Those spectroscopic parameters that we want to determine must be considered as free parameters in the computed grid of models. If some of them are kept fixed in the modeling process or the creation of the grid, one must evaluate in detail the consequences it has for the specific quantitative spectroscopic analysis to be performed.
 \item Always check carefully the various model atoms considered as input for the computations (specially in the case of those elements that will be included in the chemical abundance analysis, but also when for those elements/lines which are used to constrain the effective temperature).
 \item The step size for the various free parameters considered in the grid of models must be appropriately suited for the accuracy we want to reach in the analysis process.
\end{itemize}

The best way to acquire the necessary skills and confidence to go through all these points is to learn from someone with previous expertise or from those papers explaining the adapted strategy depending on the stars under study. Some examples of the later are provided along the next sections.

\subsection{Spectroscopic parameters}
\label{subsec:3.6}

From here onwards, things apparently become a bit more straightforward from a practical point of view. However, only expertise and a detailed and careful management of the techniques described below will allow to extract reliable information from the quantitative spectroscopic analysis to be performed. 

Once the observed spectrum is ready to be analysed (Sects.~\ref{subsec:3.2} and \ref{subsec:3.3}), and the line-broadening parameters have been determined (Sect.~\ref{subsec:3.4}), the next step is the determination of the so-called spectroscopic parameters using a suitable grid of stellar atmosphere models (Sect.~\ref{subsec:3.5.2}). In the case of the analysis of optical spectra of OB stars, these basically include the effective temperature (\Teff), the surface gravity (\grav), and the wind-strength $Q$-parameter\footnote{log~$Q$\,=\,log\,\mdot\,--\,1.5~log~$R$\,--\,1.5~log~\vinfty\ (\cite{Puls1996}). This parameter is used as a proxy of the wind properties in the optical analyses because this spectral window does not include any diagnostic line reacting exclusively (or mainly) to the mass-loss rate (\mdot) or the wind terminal velocity (\vinfty).}. In addition, there are other secondary parameters -- such as the helium abundance ($Y_{\rm He}$), the microturbulence (\micro), the exponent of the wind velocity law ($\beta$), and the abundance of the element whose ionizing equilibrium is used to determine the effective temperature (e.g., silicon in the early-B type stars) -- which need to be also determined at the same time during the analysis process.

While in the whole OB star domain the wings of the Balmer lines are the main diagnostic to estimate the surface gravity, the set of diagnostic lines that is used to constrain the effective temperature of the star depends on its spectral type. As indicated in Sect.~\ref{subsec:2.3} the \ion{He}{i} and \ion{He}{ii} lines have traditionally been considered for the analysis of mid and late O-type stars (e.g. \cite{Herrero1992, Repolust2004, Holgado2018}). The basics of this type of analysis is summarized in Figure~\ref{fig-7}, where an illustrative set of H and \ion{He}{i-ii} lines for the O9\,V star HD~214680 (10~Lac) is depicted. In addition to the observed spectrum, two synthetic spectra computed with {\sc fastwind} are overplotted. One of them is the best fitting model resulting from the {\sc iacob-gbat} analysis\footnote{{\sc iacob-gbat} (\cite{SimonDiaz2011}) is a grid-based automatic tool for the quantitative spectroscopic analysis of O-stars. The tool consists of an extensive grid of FASTWIND models, and a variety of programs implemented in IDL to handle the observations, perform the automatic analysis, and visualize the results. The tool provides a fast and objective way to determine the stellar parameters and the associated uncertainties of large samples of O-type stars within a reasonable computational time.} (see \cite{Holgado2018}); the second one represent a model in which the associated values for \Teff, \grav, and log~$Q$ have been shifted from the best fitting values to illustrate the effect on the various diagnostic lines. On the one hand, this figure serves to realize the quality of the fits our state-of-the-art models are reaching; on the other hand, it shows how the second model does not fit any of the lines since it has a too low gravity (wings of H$_{\beta}$ less extended than in the observed spectrum), a too high effective temperature (the \ion{He}{ii}\,4541 line in the model is too strong), and a too high value of the wind-strength $Q$-parameter (H$_{\alpha}$ is in emission in the model while it is not in the observed spectrum). 

\begin{figure}[t!]
\begin{center}
\includegraphics[width=0.32\textwidth, angle=90]{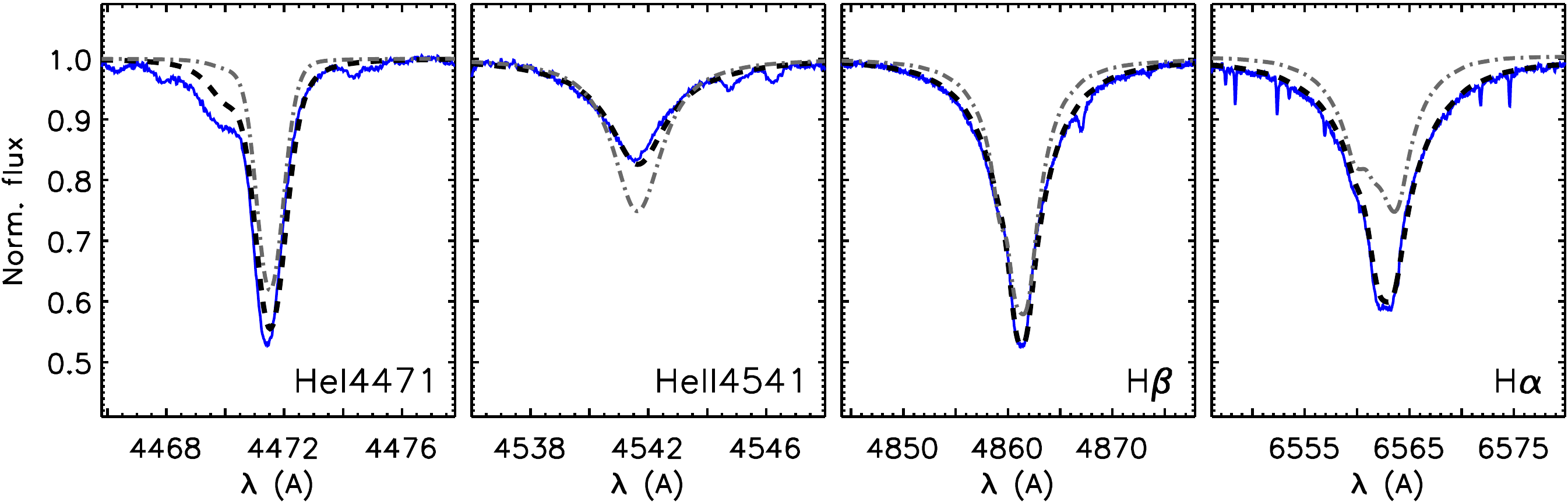}
\caption{Solid blue line: Four representative H and \ion{He}{i-ii} lines of the observed spectrum of the O9~V star HD~214680. Dashed and dashed-dotted lines: Two synthetic spectra computed with the {\sc fastwind} stellar atmosphere code, corresponding to the best fitting model resulting from the {\sc iacob-gbat} analysis (\Teff\,=\,35~000\,K, \grav\,=\,3.9\,dex, log~$Q$\,=\,-14.0) of the observed spectrum, and a model with \Teff\,=\,37000\,K, \grav\,=\,3.7, and log~$Q$\,=\,-12.5, respectively.}
\label{fig-7}       
\end{center}
\end{figure}

The basics of the strategy followed for the determination of the spectroscopic parameters in O-type stars can be easily understood with the simple example above. However, it is important to note that the situation is a bit more complex since, actually, there is not an unique, separated dependence of the various diagnostic lines with the different parameters. For example, an increase in \grav\ for a given \Teff\ produces -- in addition to a more pronounce Stark broadening of the Balmer lines due to the larger electron density in the photosphere --, weaker \ion{He}{ii} lines and stronger \ion{He}{i} lines. The larger electron density favours the recombination of higher ions into lower ionization stages, hence increasing the relative population of lower ions with respect to the higher ones. Since the intensity of \ion{He}{i} lines depend on the number He$^{+}$ ions (these are recombination lines), a larger surface gravity produces stronger \ion{He}{i} lines (and opposite for \ion{He}{ii} lines). Eventually, this implies that a model with a larger \Teff\ is required to recover the same ratios of \ion{He}{ii} to \ion{He}{i} lines when compared to a model with a lower surface gravity. In a similar way, a model with a larger value of log~$Q$ (needed, e.g., to fit a H$_{\alpha}$ line in emission) will require a larger value of \grav\ (w.r.t. a model with a weaker wind) to fit the wings of the other Balmer lines. Therefore, in the final interpretation of the outcome of any type of quantitative spectroscopic analysis (not only in O-type stars, but also in the B star domain), it is important to remember that there exist important covariances between some of the spectroscopic parameters.

In early O-type stars (O2 and O3), the \ion{He}{i} lines become too weak and the determination of the effective temperature is hence based on \ion{N}{iv-v} lines (e.g. \cite{RiveroGonzalez2012a, RiveroGonzalez2012b}). Similarly, the \ion{He}{ii} lines are absent in B-type stars, where \ion{Si}{iv-iii} and/or \ion{Si}{ii-iii} are utilized instead (e.g. \cite{McErlean1999, Urbaneja2005b, Lefever2007, SimonDiaz2010}). Figure~\ref{fig-8} shows the behavior of the equivalent widths of several diagnostic lines of N and Si (in addition to He) which are commonly used to determine temperatures in OB stars. From inspection of this figure one can easily understand why different line ratios are needed depending on the specific range in \Teff. Note, however, that I only represent the dependence of the equivalent width of the various lines with \Teff, while some of these line may also present some dependences with other parameters. Note that, whenever possible (as is always the case in the spectral type range O2\,--\,B3), the ratio of equivalent widths of two ions from the same element should be the preferred diagnostic. In this way, we eliminate the dependence of the line ratio with the abundance of the considered element.     

\begin{figure}[t!]
\begin{center}
\includegraphics[width=0.48\textwidth, angle=90]{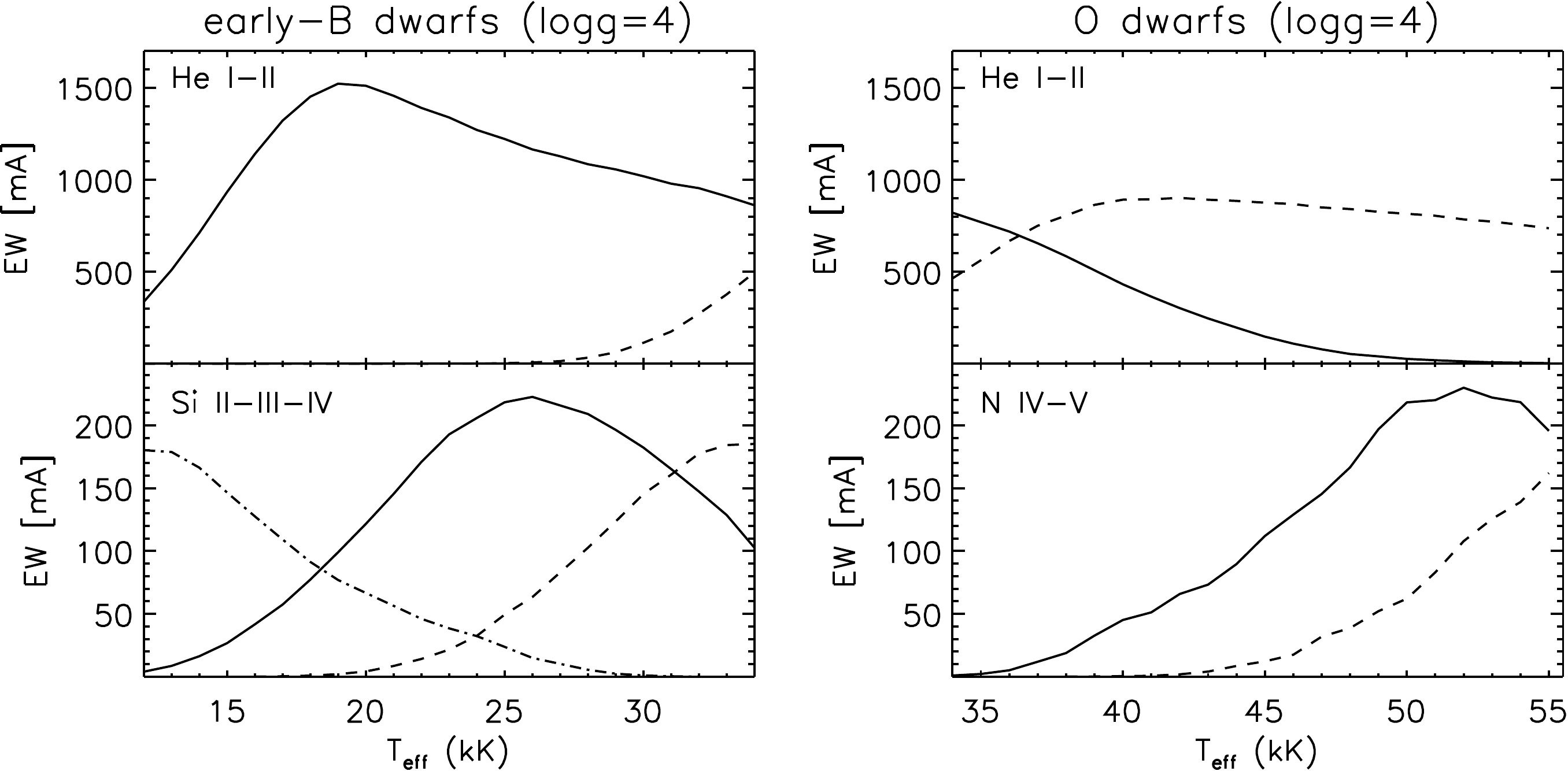}
\caption{Top panels: Behaviour of the equivalent width of the \ion{He}{i}$\lambda$4471\,\AA\ and \ion{He}{ii}$\lambda$4541\,\AA\ lines with \Teff\ in the early-B and O dwarf domain. As illustrated by the figures, the \ion{He}{i-ii} ionization balance cannot be used below \Teff\,$\sim$\,30\,kK and above 47\,kK. Bottom panels: Alternative diagnostic lines used to constraint \Teff\ in early-B (\ion{Si}{ii-iv}) and O stars (\ion{N}{iv-v}), respectively. In this specific case, the figures depicts the behaviour with \Teff\ of the following lines: \ion{Si}{ii}$\lambda$4128\,\AA\, \ion{Si}{iii}$\lambda$4552\,\AA\, \ion{Si}{iv}$\lambda$4116\,\AA\, \ion{N}{iv}$\lambda$4058\,\AA\, and \ion{N}{v}$\lambda$4603\,\AA. Note: the equivalent width of all considered lines have been obtained from a grid of {\sc fastwind} models at solar metallicity computed by the author.}
\label{fig-8}       
\end{center}
\end{figure}

I stop here due to space limitations of the chapter, but some further reading to deepen in this part of the quantitative spectroscopic analysis can be found in any of the references quoted along this subsection (see also Sects.~\ref{sec:3}). I also refer to \cite{Nieva2013} for an interesting discussion about why the spectroscopic approach to determine effective temperatures and surface gravities in early-B stars should be always preferred to the use of photometric indices, a common practice in the past, when we did not have the adequate tools to perform a proper quantitative spectroscopic analysis available.

\subsection{Photospheric abundances}
\label{subsec:3.7}

There are two different approaches for the abundance analysis in stellar objects: the {\em curve of growth} method and the {\em spectral synthesis} method. The curve of growth method is based on the behaviour of the line strength with an increase in the chemical abundance, also incorporating the effect of microturbulence\footnote{Microturbulence ($\xi_{\rm t}$) is a free parameter that was included in the stellar abundance analyses to solve the discrepancy found in the line abundances from weak and strong lines. Its physical meaning is supposed to be related to the small scale turbulent motions of the stellar plasma which could mainly affect the strong lines close to saturation.}. This method uses line equivalent widths, and hence does not require any knowledge of the exact rotational and macroturbulent broadening mechanisms affecting the line profiles. 

Figure~\ref{fig-9} provides a quick overview of the various steps followed in the oxygen abundance analysis of a narrow line early-B type star by means of the curve of growth method. I always recommend to any novice in the business to start by inspecting in detail what is summarized in the figure or, even better, perform the analysis of a similar star from scratch. The main reason is that the optical spectra of narrow line, early-B type stars include a lot of isolated \ion{O}{ii} and \ion{Si}{iv-iii} (or \ion{Si}{iii-ii} lines). Hence, performing such an exercise allows the new spectroscopist to understand all the critical points which can affect the outcome of any abundance analysis using a well behave case, before jumping to less optimal cases in which, e.g., there is one or two available lines, or those critical cases cannot be easily identified given the adopted strategy\footnote{This is the case for the spectral synthesis method, where the effect of microturbulence or the existence of wrongly modeled lines (see \cite{SimonDiaz2010}) is not so easily identified.}.

\begin{figure}[t!]
\begin{center}
\includegraphics[width=\textwidth]{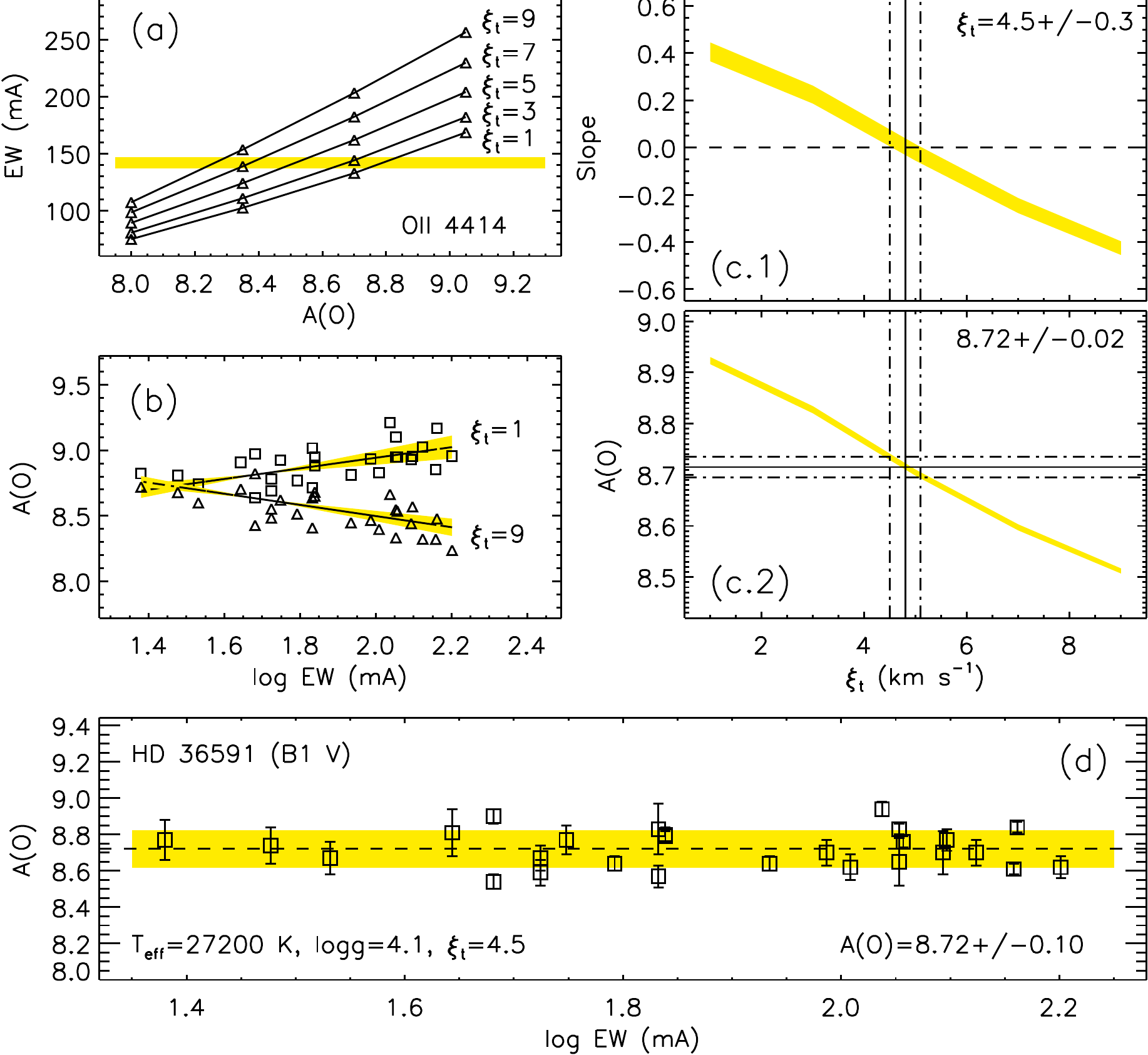}
\caption{Visual summary of the various steps comprising the oxygen abundance determination -- via the curve of growth method -- of the B1~V star HD~36591. Up to 40 \ion{O}{ii} lines are available in the optical spectrum (mainly between 4000 and 5100\,\AA), but only 27 of them have been finally considered as reliable. See text for explanation, and more details about the observed spectrum considered and the analysis process in \cite{SimonDiaz2010}.}
\label{fig-9}       
\end{center}
\end{figure}

In the curve of growth method, once the stellar parameters have been established, a grid of stellar atmosphere models whereby the abundance for the studied element and the microturbulence are varied (the remaining parameters are kept fixed) is computed. In this way, the curves of growth for each line can be constructed  by plotting the theoretical equivalent width for each value of $\xi_{\rm t}$ as a function of abundance (see Fig.~\ref{fig-9}a). From the observed equivalent width and its error, an abundance (and its uncertainty) can be derived for each line and each value of $\xi_{\rm t}$. The individual line abundances are dependent on the microturbulence which affects more the strong lines than weak lines. Figure~\ref{fig-9}b shows the $A$(O)\,--\,log$EW$ diagrams for two different values of $\xi_{\rm t}$. The value of $\xi_{\rm t}$ that minimises the dependence of the line abundances on he line stregth in the $A$(O)\,--\,log$EW$ diagrams ({\em i.e.} produces a zero slope will be the adopted microturbulence). Figures~\ref{fig-9}c.1 and \ref{fig-9}c.2 show the dependence of the slope of the $A$(O) {\em vs} log$EW$ relation and of $A$(O) on $\xi_{\rm t}$. In the last step, abundance values for each line as well as their uncertainties are calculatd for the adopted microturbulence (Fig.~\ref{fig-9}d. The final abundance value is estimated through a weighted mean of the linear individual line abundance.

The curve of growth method, as described above, also allows a straightforward computation of the final uncertainty taking into account three different sources of errors: those associated with the line-to-line abundance dispersion, those derived from the error in the determined microturbulence and, finally, those referred to the uncertainties in the stellar parameters. In addition, diagrams as the one depicted in Fig.~\ref{fig-9}d can be used as a powerful diagnostic tool to check the reliability of the various lines available for the abundance determination.

The applicability of the curve of growth method is limited to those cases when the equivalent widths of individual lines can be measured. When the projected rotational velocity of the star is high ({\em i.e.} fast rotators), or when the spectral resolution is not good enough for resolving individual lines ({\em e.g.} in extragalactic studies beyond the Local Group), a different approach must be considered: the spectral synthesis method, one of the few techniques that can be applied when blending is severe. This method is based on the computation of a grid of synthetic spectra including all the observed lines, which is then directly compared to the observed spectrum to find the best fitting model. Basically, this method follows a very similar strategy as the one illustrated in Figure~\ref{fig-7}, but including many other diagnostic lines for those elements under study. In contrast to the curve of growth method, this method requires a correct broadening of the line profiles and making sure that all the elements whose lines are present in the blending are included in the line formation calculation. In addition, the final results are more sensitive to other subtleties such as, e.g., a correct radial velocity correction of the observed spectrum. 

As I said, my recommendation is to always start the learning process with a benchmark case as the one presented in Fig.~\ref{fig-9}. That way, the new spectroscopits will consolidate a strong critical sense to avoid misinterpretations of results in those cases in which the number of diagnostic lines is more limited or the quality of the observed spectrum is worse. For example, when there is only 1 or 2 lines available -- as is, e.g., the case of nitrogen in O-type stars, or magnesium in B-type stars --, the determination of the microturbulence is more critical and one will have to make a decision on the value to use.

Some illustrative examples of studies following different types of strategies for the chemical abundance analysis of OB stars can be found in the references quoted in Sect.~\ref{sec:3} or in, e.g., \cite{Daflon2001a, Daflon2001b, Urbaneja2003, Urbaneja2005b, Daflon2004, Dufton2005, Morel2006, Morel2008, Nieva2008, Nieva2011b, Martins2015b, Martins2016, Martins2017a, Martins2017b}.

\subsection{The comparison metric: from visual fitting to PCA and MCMC}
\label{subsec:3.9}

We have seen in Sects.~\ref{subsec:3.6} and \ref{subsec:3.7} that the process of determination of spectroscopic parameters and abundances via quantitative stellar spectroscopy basically consist of finding the synthetic spectrum computed with a stellar atmosphere code which result in the best possible fit to an observed spectrum\footnote{either directly or by using equivalent widths of a selected sample of diagnostic lines.}. The basic idea is simple, but two important questions quickly pop up as soon as one wants to provide results from the analysis:

\begin{enumerate}
 \item What defines the best possible fit to the observed spectrum?
 \item Is the solution unique, or we can reach a similarly acceptable solution with different combinations of stellar parameters and/or abundances?
\end{enumerate}

I am sure that the reader is presently in a good position to assert with confidence that, given the multidimensionality of the parameter space considered during the modeling process, the existence of significant covariances between parameters\footnote{f.e., effective temperature and surface gravity, abundance and microturbulence, mass loss rate and the $\beta$ parameter.} and taking into account some technical limitations related to the quality of the observed spectrum (in terms, e.g, of signal-to-noise ratio), the answer to the second question is ''no, the solution is not unique``. Indeed, a proper identification of the range of acceptable parameters/abundances (i.e. definition of the associated uncertainties) is as important as the determination of the best fitting or central values.

Regarding the first question, the considered strategy to define the central values and the associated uncertainties has gained in complexity and robustness in the last decades. Not so long ago, given the computational limitations, most of the spectroscopic analyses were made based on small grids of stellar atmosphere models, and the determination of the final solution was a subjective by eye decision, sometimes supported by some more quantitative (but still simple) arguments (e.g., \cite{Herrero1992, Villamariz2005})}. However, the  continuously  increasing  amount  of  high-quality  spectroscopic observations of massive OB stars provided by different surveys during the first decade of the XXI century (e.g., \cite{Evans2005, Evans2011, Barba2010, Barba2017, SimonDiaz2015, Wade2016}) made it clear the necessity to develop  more objective, semi-automatized techniques which allow for the extraction of information about stellar parameters and abundances (and the associated uncertainties) from large spectroscopic datasets in a reasonable computational time. 

Some notes on various of the techniques proposed to date can be found in \cite{Mokiem2005, Lefever2007, Urbaneja2008,  SimonDiaz2011, Castro2012}, and \cite{Irrgang2014}. Most of them are based on specific grids of pre-computed models and a $\chi^2$ algorithm which allow to find the best fitting solution (or central values for each of the considered free parameters) and the associated uncertainties. However, the use of other strategies based on, e.g., projection and/or pattern recognition methods (in contrast to the minimum distance methods, as the $\chi^2$ algorithms) are slowly but surely started to be explored. In addition, some works are already exploiting strategies based on the application of Genetic Algorithms (GA), principal component analysis (PCA), Gaussian process regression and Monte Carlo Markov Chains (MCMC) techniques. In particular, the later three are certainly envisaged as a promising way to minimize the computational time needed to create optimal grids of models, and to speed up the process of exploration of the multidimensional parameter space.

\begin{acknowledgement}
I want to warmly thank all those friends and colleagues from which I've been able to learn and discuss about quantitative spectroscopy since my first years as PhD student at the Instituto de Astrof\'isica de Canarias. Special thanks to A.~Herrero, M.~A.~Urbaneja, C.~Villamariz, F.~Najarro, C.~Trundle, D.~J.~Lennon, J.~Puls, N.~Castro, M.~Garcia, F.~Nieva, C. Sab\'in-Sanjulian, K. R\"ubke, G.~Holgado, S.~Berlanas, and A.~de Burgos. Let this text serves to spread part of the knowledge I've acquired from all of you during these years.  
\end{acknowledgement}

\bibliographystyle{spphys}
\bibliography{SimonDiaz}

\end{document}